\documentstyle[12pt]{article}
\input{epsf}
\newcommand{\be}{\begin{eqnarray}}
\newcommand{\ee}{\end{eqnarray}}
\newcommand{\ba}{\begin{array}}
\newcommand{\ea}{\end{array}}
\newcommand{\partialslash}{\partial\hspace{-.5em}/\hspace{.15em}}
\renewcommand{\thefootnote}{\fnsymbol{footnote}}
\begin{document}
\rightline{RUB-TPII-13/97}
\rightline{hep-ph/9712472}

\begin{center}
{\LARGE\bf Nucleon parton distributions in the large $N_c$
limit\footnote{Talk given at the Workshop ``Deep
Inelastic Scattering off Polarized Targets: Theory Meets Experiment'',
DESY--Zeuthen,  Sep.\ 1--5, 1997.}}

\vspace{1cm}
C. Weiss\footnote{E-mail: weiss@hadron.tp2.ruhr-uni-bochum.de}

\vspace*{1cm}
{\it Institut f\"ur Theoretische Physik II,
Ruhr--Universit\"at Bochum, D--44780 Bochum, Germany}\\

\vspace*{2cm}

\end{center}

\begin{abstract}
We review recent progress in calculating the twist--2 parton distribution
functions (PDF's) at a low normalization point in the 
large--$N_c$ limit, where the nucleon can be described as a soliton of the 
effective chiral lagrangian. This field--theoretic approach preserves 
all general requirements on the PDF's (positivity, normalization {\em etc}.).
In particular, it allows to calculate the polarized
and unpolarized antiquark distributions at the low normalization point.
\end{abstract}
\setcounter{footnote}{0}
\renewcommand{\thefootnote}{\arabic{footnote}}
\par
In this talk we review recent attempts to calculate the leading--twist 
parton distributions of the nucleon at a low normalization point in 
the large--$N_c$ limit \cite{DPPPW96,DPPPW97,PP96}. The results presented
here have been obtained in collaboration with D.I.\ Diakonov,
V.Yu.\ Petrov, P.V.\ Pobylitsa, and M.V.\ Polyakov.
\par
\underline{Introduction.}
The parton distribution functions (PDF's) of the nucleon are an essential
ingredient in the description of deep--inelastic scattering and a variety 
of other hard processes. The scale dependence of the PDF's in the asymptotic 
region is determined by the renormalization group equation of perturbative
QCD. A complete description of the experiments requires non-perturbative 
information in the form of the values of the parton distributions at some 
initial normalization point. Several sets of input distributions have been 
determined by parametrizing data from a number of different 
experiments at large $q^2$ \cite{MRS95,CTEQ95,GRV95,GRSV96}. 
All modern parametrizations involve both antiquarks and gluons at 
the low scale. 
\par
When attempting to compute the parton distributions at a low normalization
point from first principles one must deal with non-perturbative effects
such as the dynamical breaking of chiral symmetry and confinement.
While a satisfactory theory of confinement is still lacking, the 
dynamical breaking of chiral symmetry in QCD is well understood 
theoretically: the resulting effective theory at low energies is known, 
and a dynamical picture of chiral symmetry breaking is provided
by the instanton vacuum\footnote{For a recent review of the instanton
vacuum see refs.\cite{SchSh96,D96_Varenna}.} \cite{DP84,DP86}, which has 
recently received direct support from lattice simulations 
\cite{lattice}. A description of 
the nucleon based on the dynamical breaking of chiral symmetry
is possible in the large--$N_c$ limit, where QCD becomes equivalent to an 
effective theory of mesons, and the nucleon emerges as a soliton of the
pion field \cite{Witten,ANW,DPP88}. This picture of the nucleon is known to 
give a good account of hadronic observables such as the nucleon mass, 
magnetic moments, form factors {\em etc.} \cite{Review}. It is thus natural to
describe also the parton distribution at a low normalization point in
this approach. An important advantage of this description of the nucleon 
is its field--theoretic character, which is essential for the parton 
distributions to satisfy general requirements such as 
positivity, proper normalization {\em etc}.
\par
\underline{Effective chiral theory.}
In the large--$N_c$ limit, the consequences of the dynamical breaking of 
chiral symmetry can be encoded in an effective lagrangian for the pion field,
valid at low energies. It can be expressed as a functional integral over 
quark fields ($SU(2)$ flavor) in the background pion field, 
$\pi (x)$, \cite{DP86,DE,DSW}
\be
\exp\left( i S_{\rm eff}[\pi (x) ] \right) &=&
\int D\psi D\bar\psi \; \exp\left[ i\int d^4 x\,
\bar\psi(i\partialslash - M U^{\gamma_5})\psi\right] ,
\label{effective_action} \\
U^{\gamma_5}(x) &=& \exp\left[ i\pi^a(x)\tau^a \gamma_5 \right] .
\label{FI}
\ee
Here, $M$ is the dynamical mass (generally speaking, it is 
momentum--dependent), which is due to the 
spontaneous breaking of chiral symmetry. Eq.(\ref{effective_action}) 
describes the minimal chirally invariant interaction of quarks with Goldstone 
bosons. In the long--wavelength limit, expanding in derivatives of the 
pion field, the effective action Eq.(\ref{effective_action}) reproduces
the Gasser--Leutwyler lagrangian with correct coefficients, including the
Wess--Zumino term. The effective theory defined by 
Eq.(\ref{effective_action}) is valid for momenta up to an UV 
cutoff, which is the scale at which the dynamical quark mass drops to 
zero. For simplicity we shall take in the discussion here the quark mass
to be momentum--independent and assume divergent quantities to be made 
finite by applying some regularization scheme; below we shall see why, 
and under which conditions, this approximation is justified when 
computing (anti--) quark DF's.
\par
The effective action Eq.(\ref{effective_action}) has been derived from 
the instanton vacuum, which provides a natural mechanism of dynamical chiral
symmetry breaking and enables one to express the parameters
intrinsic in Eq.(\ref{effective_action}) --- the dynamical mass, $M$, 
and the ultraviolet cutoff --- in terms of the QCD scale parameter, 
$\Lambda_{QCD}$ \cite{DP86}. The cutoff
is given by the average instanton size,
\be
\bar\rho^{-1} &\simeq& 600 \, {\rm MeV} ,
\ee
which is the order of the scale at which the coupling constant is fixed 
in the instanton vacuum. Since the effective theory is not valid
beyond $\bar\rho^{-1}$, when applying it to compute PDF's we are certainly
dealing with distributions normalized at a scale not higher than
$600 \, {\rm MeV}$.
\par
An important point to emphasize is that, when working with the effective 
action, Eq.(\ref{effective_action}), it is implied that the ratio of the 
dynamical quark mass to the ultraviolet cutoff is parametrically small, 
$M\bar\rho \ll 1$. In the instanton vacuum this is guaranteed by 
the fact that $M^2$ is proportional to the instanton density, so that
$(M\bar\rho )^2$ is proportional to the packing fraction of the instanton 
medium,
\be
(M\bar\rho )^2 &\propto& \left( \frac{\bar\rho}{R} \right)^4 ,
\ee
which is a small parameter. However, this requirement is really of a more
general nature --- otherwise degrees of freedom other than 
those included in Eq.(\ref{effective_action}) would play an essential role
in the effective dynamics, and the concept of an effective action would
not be very meaningful. One may interpret the inverse ultraviolet 
cutoff, $\bar\rho$, as the size of the ``constituent'' quark described by
the effective theory. The smallness of $M\bar\rho$ implies that
the constituent quark is nearly pointlike --- its size is small
compared to its typical wavelength in the nucleon, which is of 
order\footnote{The dominant contribution to the twist--2 quark and antiquark
DF's comes from quarks with momenta of order $M \ll \bar\rho^{-1}$, 
see below.} $M^{-1}$.
This fact is essential for the interpretation of the distribution 
functions computed with the effective chiral theory.
For $M\bar\rho \rightarrow 0$ one may identify them with the QCD quark 
and antiquark distributions. When trying to be more accurate, keeping
terms of order $(M\bar\rho )^2$, one begins to resolve the structure of the 
``constituent'' quark, and the distribution functions computed in the
effective theory should be regarded as distributions of 
composite objects which have a substructure in terms of QCD partons.
\par
To describe the gluon distribution one needs a microscopic picture
of the non-perturbative gluon degrees of freedom. Such a picture
is provided by the instanton vacuum, which on one hand
allows to derive the effective chiral action, Eq.(\ref{effective_action}),
by explicit integration over gluon degrees of freedom, on the other 
hand makes it possible to directly evaluate matrix elements of gluon
operators, using the effective operator method developed in
ref.\cite{DPW95}. One finds that the twist--2 gluon distribution is of
order $(\bar\rho /R )^4$, which is consistent with the above
statement that the structure of the constituent quark starts
to appear at level $(M\bar\rho )^2$. To compute the twist--2 gluon 
distribution one needs to construct both the effective action
and the effective operators at order $(M\bar\rho )^2$.
In the computation of twist--2 quark and antiquark DF's reported here 
we work in the limit $M\bar\rho \rightarrow 0$, where the constituent
quark is pointlike and the twist--2 gluon distribution is 
zero. For a more detailed discussion of the gluon 
distribution, including the role of the gauge field in the twist--2
quark operators, see refs.\cite{BPW97,PW97}.\footnote{
Calculations of the twist--2 gluon distribution in the instanton vacuum
using a different approach have recently been performed by
Kochelev \cite{Kochelev}. However, the
contributions taken into account there do not represent the
full answer to order $(\bar\rho / R)^4$; see refs.\cite{BPW97,PW97} for a 
discussion.}
\par
\underline{The nucleon as a chiral soliton.}
In the effective theory defined by Eq.(\ref{effective_action}) the
nucleon is in the large--$N_c$ limit described by a static classical pion
field (``soliton'') \cite{DPP88}. In the nucleon rest frame it is 
of ``hedgehog'' form,
\be
U^{\gamma_5}_c ({\bf x}) &=& 
\exp\left[ i n^a \tau^a P(r) \gamma_5 \right],
\hspace{1cm} r \; = \; |{\bf x}|,
\hspace{1cm} {\bf n} \; = \; \frac{{\bf x}}{r} ,
\label{hedge}
\ee
where $P(r)$ is called the profile function, satisfying $P(0) = -\pi , 
P(r) \rightarrow 0$ for $r\rightarrow\infty$, which is
determined by minimizing the energy of the static pion field. Quarks are
described by single-particle wave functions, which are the solutions
of the Dirac equation in the external pion field,
\be
H\Phi_n &=& E_n\Phi_n ,
\hspace{1.5cm}
H \;\;
= \;\; -i\gamma^0 \gamma^k \partial_k + M \gamma^0 U^{\gamma_5}_c \,.
\label{dirac}
\ee
The spectrum of the one-particle Hamiltonian, $H$, contains a discrete
bound--state level, which must be occupied by $N_c$ quarks to
have a state of unit baryon number, as well as the positive and negative
energy Dirac continuum, distorted by the presence of the pion
field. The nucleon mass is given by the
minimum of the bound--state energy and the aggregate energy of the negative
Dirac continuum, the energy of the free Dirac continuum subtracted
\cite{DPP88}. Nucleon states of definite spin-isospin and 3-momentum are 
obtained by quantizing the rotational and translational zero modes of the 
soliton, {\em i.e.}, applying a flavor rotation and a shift of the 
center to the soliton field, Eq.(\ref{hedge}), and 
integrating over the corresponding collective 
coordinates with appropriate wave functions \cite{DPP88,Review}.
\par
\underline{Quark-- and antiquark distributions from the effective chiral
theory.}
In order to compute the twist--2 quark and antiquark distribution functions 
in the effective chiral theory one may start either from their
``parton model'' definition as numbers of particles carrying a
given fraction of the nucleon momentum in the infinite--momentum frame,
or from the ``field--theoretic'' definition as forward matrix elements
of certain light--ray operators in the nucleon. Both ways lead to 
identical expressions for the quark DF's in the chiral 
soliton model. This remarkable fact should be attributed to the 
field--theoretic
nature of this description of the nucleon, and to the fact that the main 
hypothesis of the Feynman parton model --- that transverse momenta do 
not grow with $q^2$ \cite{Feynman} --- is satisfied within this model.
In the first formulation one considers a nucleon with large 
3--momentum,
\be
P_N &=& \frac{M_Nv}{\sqrt{1-v^2}},
\hspace{1.5em} v \; \rightarrow \; 1 ,
\ee
and defines the quark and antiquark DF's as
\be
\left. 
\ba{l} D_{i, f} (x) \\[.5cm] \bar D_{i, f} (x) \ea \right\}
&=&
\int\frac{d^3k}{(2\pi)^3}2\pi \delta\left( x - \frac{k^3}{P_N}\right)
\langle N_{{\bf v}}| \;\;
\left\{ \ba{l}
a_{i, f}^+ ({\bf k}) a_{i, f} ({\bf k}) \\[.5cm]
b_{i, f}^+ ({\bf k}) b_{i, f} ({\bf k}) \ea \right\} \;\;
|N_{{\bf v}}\rangle ,
\label{distributions}
\ee
where $|N_v \rangle$ denotes the state with the fast--moving nucleon
and $a_{i, f}^+, a_{i, f}$ and $b_{i, f}^+, b_{i, f}$ are the 
creation/annihilation operators for quarks (antiquarks), characterized by 
a polarization quantum number $i$ and flavor $f$. For large $N_c$
the nucleon consists of $O(N_c )$ quarks and antiquarks, so we can hope
to describe the distribution functions for values of $x$ of order
\be
x &\sim& \frac{1}{N_c}. 
\ee
In the large $N_c$ limit the nucleon matrix element Eq.(\ref{distributions})
reduces to a sum of matrix elements between quark single--particle 
states in the mean pion field (saddle point approximation), which is
given in this case by the hedgehog field, Eq.(\ref{hedge}), 
boosted to velocity $v$, see \cite{DPPPW97} for details. 
Taking the limit $v \rightarrow 1$ one arrives at the following 
expressions for the twist--2 quark and antiquark distributions of the 
nucleon in the large--$N_c$ limit:
\be
\lefteqn{
\left. 
\ba{l} D_{i, f} (x) \\[.5cm] \bar D_{i, f} (x) \ea \right\}
} && \nonumber \\
&=& N_c M_N 
\left\{
\ba{c} {\displaystyle \sum_{\rm occ.}} \\[.5cm] 
{\displaystyle \sum_{\rm non-occ.}}
\ea \right\}
\int\!\frac{d^3k}{(2\pi)^3}
\Phi_{n, f}^\dagger ({\bf k}) (1+\gamma^0\gamma^3) \gamma^0 \Gamma_i
\delta (k^3 + E_n \mp xM_N) \Phi_{n, f} ({\bf k}) ,
\nonumber \\ 
\label{FinalQuarks}
\ee
where, for example, 
\be
\Gamma_i &=& \gamma^0 \frac{1 \pm \gamma_5}{2}
\ee
for quarks polarized along or against the direction 
of the nucleon velocity. Here the sums run over all occupied or 
non-occupied quark levels in the 
hedgehog pion field, that is, the bound--state level and the negative 
Dirac continuum or the positive Dirac continuum, respectively.
It is understood in Eq.(\ref{FinalQuarks}) that one subtacts
the corresponding sum over eigenstates of the free Hamiltonian,
where $\pi (x) = 0$. 
\par
Alternatively, one can start from the QCD definitions of 
the twist--2 quark distribution functions as matrix elements of
certain light--ray operator, which act as generating functions of the
series of local twist--2 operators \cite{Collins-Soper-82,Jaffe95}; 
in the unpolarized case
\be
D_{{\rm unpol}, f} (x) &=&  \frac{1}{4\pi}
\int\limits_{-\infty}^\infty dz^- \, e^{ixp^+z^-} \,
\langle P | \, \bar\psi_f (0) \, \gamma^+ \, \psi_f (z) \, | P \rangle 
\Bigr|_{z^+=0,\>z_\perp=0}\, , \nonumber \\
\bar D_{{\rm unpol}, f} (x) &=& - \left\{ x \rightarrow -x \right\} ,
\label{nonlocal}
\ee
and in the longitudinally polarized case
\be
D_{{\rm pol}, f} (x) &=&  \frac{1}{4\pi}
\int\limits_{-\infty}^\infty dz^- \, e^{ixp^+z^-} \,
\langle P, S | \, \bar\psi_f (0) \, \gamma^+  \gamma_5 \, 
\psi_f (z) \, | P, S \rangle 
\Bigr|_{z^+=0,\>z_\perp=0}\, , \nonumber \\
\bar D_{{\rm pol}, f} (x) &=& \left\{ x \rightarrow -x \right\} .
\label{nonlocal_pol}
\ee
These matrix elements can be evaluated in the nucleon rest frame,
expanding the quark fields in Eqs.(\ref{nonlocal}, \ref{nonlocal_pol}) 
in the basis of single--particle wave functions, Eq.(\ref{dirac});
the result is identical to Eq.(\ref{FinalQuarks}).
\par
To obtain the unpolarized or polarized (anti--) quark distributions
for a nucleon state of definite spin and isospin, one has to
apply a flavor rotation to the quark wave functions in the basic 
expression Eq.(\ref{FinalQuarks}), 
$\Phi_{n, f} \rightarrow R_{ff'}(t) \Phi_{n, f'}$, and integrate over 
slow rotations, projecting on a nucleon state with given spin and 
isospin \cite{DPPPW96}. At this point characteristic differences between 
the different isospin combinations of the unpolarized and polarized
distributions appear. The angular velocity of the soliton, 
$R^\dagger (\partial R / \partial t)$, is of order $1/N_c$
(the soliton moment of inertia is $O(N_c)$), and in the leading order
of the $1/N_c$--expansion one may neglect it. The distribution functions
which are non-zero at this level are the isosinglet unpolarized and
isovector polarized --- they are the leading ones in the $1/N_c$--expansion.
The isovector unpolarized and isosinglet polarized distributions,
on the other hand, are non-zero only after expanding to first
order in the angular velocity; they are thus suppressed relative to
the former by a factor $1/N_c$. Thus the $1/N_c$--expansion leads
to the following classification of quark/antiquark DF's in
``large'' and ``small'' ones:
\begin{center}
\begin{tabular}{|l|c|c|c|}
\hline
     & unpolarized & polarized (longitud.) & polarized (transv.) \\ 
\hline
``large'' & $u + d$ & $\Delta u - \Delta d$ & $h_1^u - h_1^d$  \\
\hline
``small'' & $u - d$ & $\Delta u + \Delta d$ & $h_1^u + h_1^d$  \\
\hline 
\end{tabular}
\end{center}
More precisely, bearing in mind that $x\sim 1/N_c$, we can say that
the (anti--) quark DF's in the large $N_c$ limit behave as
\be
D^{\rm large}(x) &\sim& N_c^2 \, f(N_c x) , \hspace{2cm}
D^{\rm small}(x) \;\; \sim \;\; N_c \, f(N_c x) ,
\ee
where $f(y)$ is a stable function in the large $N_c$--limit which
depends on the particular distribution considered. One may easily
convince oneself that this is consistent with
the well--known $N_c$--dependence of the lowest moments of the DF's:
the first moments of $u + d$ (baryon number) and of $\Delta u - \Delta d$ 
(isovector axial coupling, $g_A^{(3)}$) are $O(N_c)$, while 
those of $u - d$ (isospin) and of $\Delta u + \Delta d$ 
(isosinglet axial coupling, $g_A^{(0)}$) are $O(N_c^0)$.
\par
The ``large'' DF's are given by simple sums over quark 
levels; the isosinglet unpolarized DF by (the sum over flavor indices
is understood on the R.H.S)
\be
[u + d] (x) &=& N_c M_N {\displaystyle \sum_{\rm occ.}}
\int\!\frac{d^3k}{(2\pi)^3}
\Phi_n^\dagger ({\bf k})(1+\gamma^0\gamma^3) \delta (k^3 + E_n + xM_N)
\Phi_n ({\bf k}) ,
\label{singlet_quark} \\ 
{} [\bar u + \bar d] (x) 
&=& N_c M_N {\displaystyle \sum_{\rm non-occ.}}
\left\{ x \rightarrow - x \right\} .
\label{singlet_anti}
\ee
An alternative representation of the DF is obtained
by applying in Eq.(\ref{nonlocal}) the anticommutation 
relation of quark fields before taking the light--cone limit; 
this leads to a representation of the quark DF's as sums over 
non-occupied, and of the antiquark DF as a sum over occupied states: 
\be
[\bar u + \bar d] (x) 
&=& - N_c M_N {\displaystyle \sum_{\rm occ.}}
\left\{ x \rightarrow - x \right\} ,
\label{singlet_anti_occ} 
\ee
where the braces again denote the corresponding expressions appearing 
in Eq.(\ref{singlet_quark}). One may regard Eq.(\ref{singlet_quark}) 
together with Eq.(\ref{singlet_anti_occ}) as one 
universal function, describing the quark DF at $x > 0$ and 
minus the antiquark distribution at $x < 0$. 
The isovector--polarized DF is given by\footnote{For the corresponding
formulas in the case of transverse polarization, see ref.\cite{PP96}.}
\be
[\Delta u - \Delta d] (x) 
&=& -\frac{1}{3} (2T_3) N_c M_N 
{\displaystyle \sum_{\rm occ.}}
\int\!\frac{d^3k}{(2\pi)^3}
\Phi_n^\dagger ({\bf k})(1+\gamma^0\gamma^3) \gamma_5 
\delta(k^3 + E_n + xM_N) \Phi_n ({\bf k}) ,
\nonumber \\
\label{isovector_quark} \\
{}[\Delta \bar u - \Delta \bar d] (x)
&=& \frac{1}{3} (2T_3) N_c M_N 
{\displaystyle \sum_{\rm non-occ.}}
\left\{ x \rightarrow -x \right\} ,
\label{isovector_anti}
\ee
or, alternatively,
\be
[\Delta \bar u - \Delta \bar d] (x) 
&=& - \frac{1}{3} (2T_3) N_c M_N 
{\displaystyle \sum_{\rm occ.}}
\left\{ x \rightarrow -x \right\} ,
\label{isovector_anti_occ}
\ee
where $2T_3 = \pm 1$ for proton and neutron, respectively. 
The ``small'' DF's in the large--$N_c$ limit are given by double 
sums over quark levels divided by the moment of inertia of the soliton; 
see ref.\cite{DPPPW96} for details.
\par
From the explicit representations 
Eqs.(\ref{singlet_quark}--\ref{isovector_anti_occ}) it can easily be seen 
that the large--$N_c$ quark and antiquark DF's satisfy the following 
properties: 
\\
{\em Positivity:} The singlet unpolarized 
quark and antiquark DF's, Eqs.(\ref{singlet_quark}, \ref{singlet_anti}), 
are explicitly positive (the Dirac matrices in these expressions are
a projector and thus positive definite). 
\\
{\em Sum rules:} Integrating Eq.(\ref{singlet_quark}) minus the
representation Eq.(\ref{singlet_anti_occ}) over $x$ one finds
\be
\int dx [u + d - \bar u - \bar d] (x)
&=& N_c \left(\mbox{number of occupied levels in the soliton} \right.
\nonumber \\
&& - \left. \mbox{number of occupied levels in the vacuum} \right) ,
\ee
which coincides with the baryon number of the nucleon in the 
large--$N_c$ limit. One can also show that the
momentum sum rule is satisfied by virtue of the fact that the
soliton field is a stationary point of the static energy.
Similarly, integrating $\Delta u - \Delta d + \Delta\bar u - \Delta\bar d$, 
Eq.(\ref{isovector_quark}) minus Eq.(\ref{isovector_anti_occ}), one obtains 
the expression for the isovector axial coupling, $g_A^{(3)}$, in the
chiral soliton model \cite{DPP88,Review}, {\em i.e.}, the Bjorken sum rule 
is satisfied within this approach.
\\
For a proof of sum rules for the ``small'' DF's
({\em e.g.}\ the isospin sum rule for $u - d$), and a discussion of 
the Gottfried sum we refer to ref.\cite{DPPPW96}.
\par
It is essential that both the contributions of the bound--state 
level as well as the Dirac continuum are taken into account in the
representation of the large--$N_c$ quark and antiquark DF's, 
Eqs.(\ref{singlet_quark}--\ref{isovector_anti_occ}).
Restricting oneself to the bound--state level only one would, for
example, violate either the the baryon number sum rule or the
positivity of the singlet unpolarized antiquark 
distribution --- since the level makes a {\em negative} contribution
to Eq.(\ref{singlet_anti_occ}), depending 
on whether one writes the antiquark distribution in the form 
Eq.(\ref{singlet_anti}) or Eq.(\ref{singlet_anti_occ}). Adding the
contribution of the Dirac continuum Eq.(\ref{singlet_anti}) and 
Eq.(\ref{singlet_anti_occ}) give identical results, the antiquark 
distribution is positive, and the baryon number unity.
\par
In the discussion of the large--$N_c$ DF's so far we have not explicitly 
taken into account the UV cutoff, which is an essential ingredient in the 
effective chiral theory, Eq.(\ref{effective_action}).
In fact, the expressions for the distribution functions as sums over 
quark levels, Eqs.(\ref{singlet_quark} -- \ref{isovector_anti_occ}) 
contain UV divergences which are made finite by the UV cutoff.  
When choosing a regularization method
to implement this UV cutoff it is of utmost importance that the
regularization do not violate any of the fundamental properties of
the DF's. A crucial requirement is that the regularization
should preserve the completeness of the set of quark single--particle
wave functions in the soliton field, Eq.(\ref{dirac}). An incomplete set 
of basis functions would mean a violation of the local anticommutation
relation of the quark fields, that is, a violation of causality.
For an extensive discussion of the many facets of this we refer 
to ref.\cite{DPPPW97}. Let us mention here only that the equivalence
of the representations of the (anti--) quark DF's as sums over
non-occupied or occupied states, {\em cf.}\ 
Eqs.(\ref{singlet_anti}, \ref{singlet_anti_occ}), which is
instrumental in ensuring {\em both} positivity and proper 
normalization of the isosinglet DF, depends crucially on the completeness 
of the basis.
\par
A regularization by subtraction, for example, a Pauli--Villars cutoff,
meets the above requirement, while a regularization by cutoff (for example,
an energy cutoff or the popular proper--time regularization of the fermion
determinant) is equivalent to working with an incomplete set of states 
and thus unacceptable. The Pauli--Villars cutoff is implemented by
subtracting from the sums 
Eqs.(\ref{singlet_quark} -- \ref{isovector_anti_occ}) a multiple
of the corresponding sums computed with a regulator mass, 
$M_{PV} \simeq \bar\rho^{-1}$,
\be
D(x)_{PV} &=& D(x)\,\, \rule[-1.0em]{.25mm}{2em}_{\, M}
 - \frac{M^2}{M_{PV}^2} D(x) \,\,\rule[-1.0em]{.25mm}{2em}_{\, M_{PV}}.
\label{PV}
\ee
It can be shown that such regularization preserves all qualitative properties 
of the DF's (positivity, sum rules) for which the completeness of states
is an essential condition, thus justifying our above discussion of
``naive'' unregularized expressions \cite{DPPPW97}. (Note that only 
UV divergent quantities such as the total isosinglet distribution
$u + d + \bar u + \bar d$ should be regularized; UV finite 
quantities such as the valence distribution $u + d - \bar u - \bar d$
should be left unregularized.) In short, the UV cutoff in the effective 
theory Eq.(\ref{effective_action}) 
can be implemented in a way that the field--theoretic character of the 
description of quark and antiquark DF's is preserved.
\par
It is important to note that, with a regularization preserving completeness,
the actual dependence of the twist--2 DF's on the cutoff of the 
effective theory is very weak: the DF's 
Eqs.(\ref{singlet_quark} -- \ref{isovector_anti_occ}) are only logarithmically
divergent for fixed $x$, and so are all their moments, see ref.\cite{DPPPW97}.
The dominant contributions to the DF's thus come from degrees of freedom
with momenta of order $M \ll \bar\rho^{-1}$. This fact is crucial for the
consistency of the effective theory approach: the quark and antiquark DF's are 
determined by genuine dynamics of the effective theory --- that is, by 
degrees of freedom with momenta much smaller than the cutoff --- not by 
the details of the UV regularization, which, at any rate, is known 
only schematically. In particular, this justifies working with
a suitable generic regularization (Pauli--Villars subtraction) rather than 
with the momentum--dependent quark mass obtained from the instanton vacuum.
\par
One exception to the above is the region of values 
of $x \le (M\bar\rho )^2 / N_c$, where the dominant contribution to the 
(anti--) quark DF's comes from states with large (positive or negative) 
energies\footnote{We remind that we are discussing the (anti--) quark 
DF's at a low normalization point, whose small--$x$ behavior has no direct
relation to that of the DF's at experimental scales, which is
determined essentially by perturbative evolution.}. 
In this parametrically small region of $x$ the momentum dependence of the
quark mass is essential and, in the case of the 
singlet unpolarized distribution, Eqs.(\ref{singlet_quark}, 
\ref{singlet_anti_occ}), leads to a 
vanishing of the distribution at $x = 0$, as has recently been shown 
in connection with an investigation of the off-forward PD's in 
the large--$N_c$ approach 
\cite{PPPBGW97}. We note also that, in contrast to the twist--2 DF's, 
matrix elements of higher--twist operators generally receive contributions
from momenta of order of the cutoff, $\bar\rho^{-1}$, when computed in the
effective theory. Such matrix elements can be estimated within the 
effective theory derived from the instanton vacuum, taking into account 
the full momentum dependence of the constituent quark 
mass \cite{BPW97,PW97}.
\par
\underline{Results for quark and antiquark distributions.}
To compute the DF's one needs to evaluate the sums over quark levels,
Eqs.(\ref{singlet_quark} -- \ref{isovector_anti_occ}), taking into account 
the contributions of the bound--state level as well as the polarized
Dirac continuum. This can be done either by numerical diagonalization 
of the hamiltonian \cite{DPPPW97}, 
or by rewriting the sums over levels in terms of the quark Green function
in the background pion field, which allows an approximate evaluation
of the continuum contribution using the so-called interpolation
formula \cite{DPPPW96}.
%
%
\begin{figure}[t]
\epsfxsize=15cm
\epsfysize=12cm
\epsffile{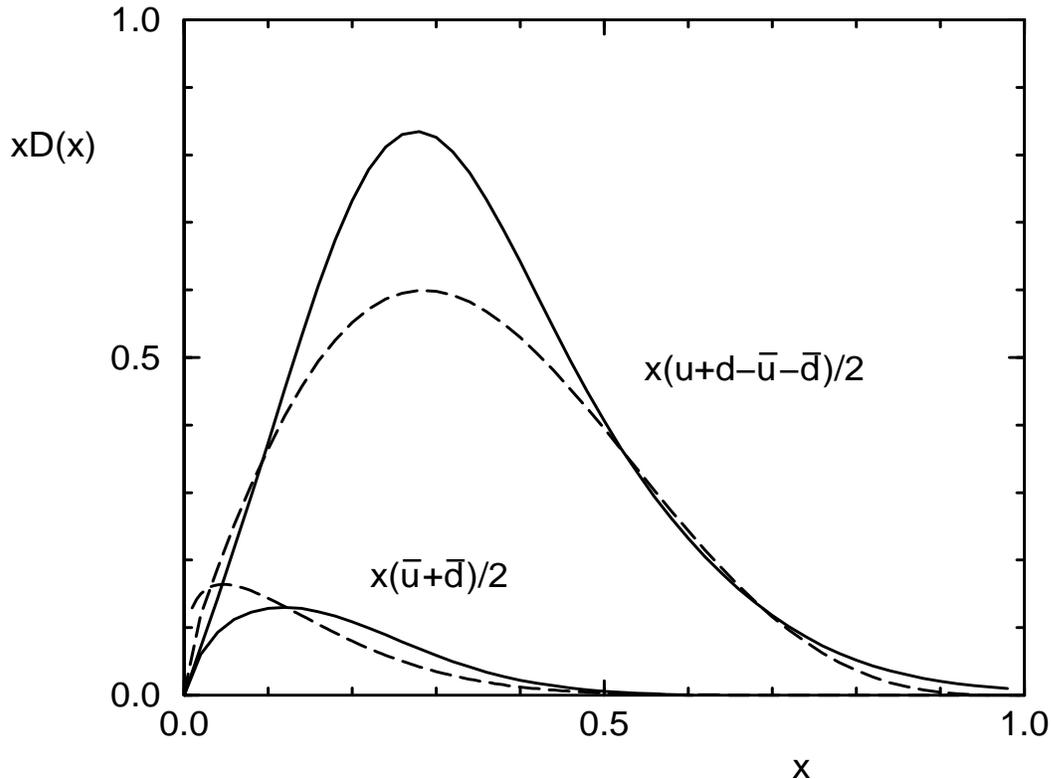}
\caption[]{{\em Solid lines}: The isosinglet unpolarized valence quark and 
antiquark distributions computed in the large--$N_c$ limit 
\cite{DPPPW96,DPPPW97}. {\em Dashed lines}: NLO--parametrization 
of GRV \cite{GRV95}.}
\label{fig_1}
\end{figure}
%
%
\begin{figure}[t]
\epsfxsize=15cm
\epsfysize=12cm
\epsffile{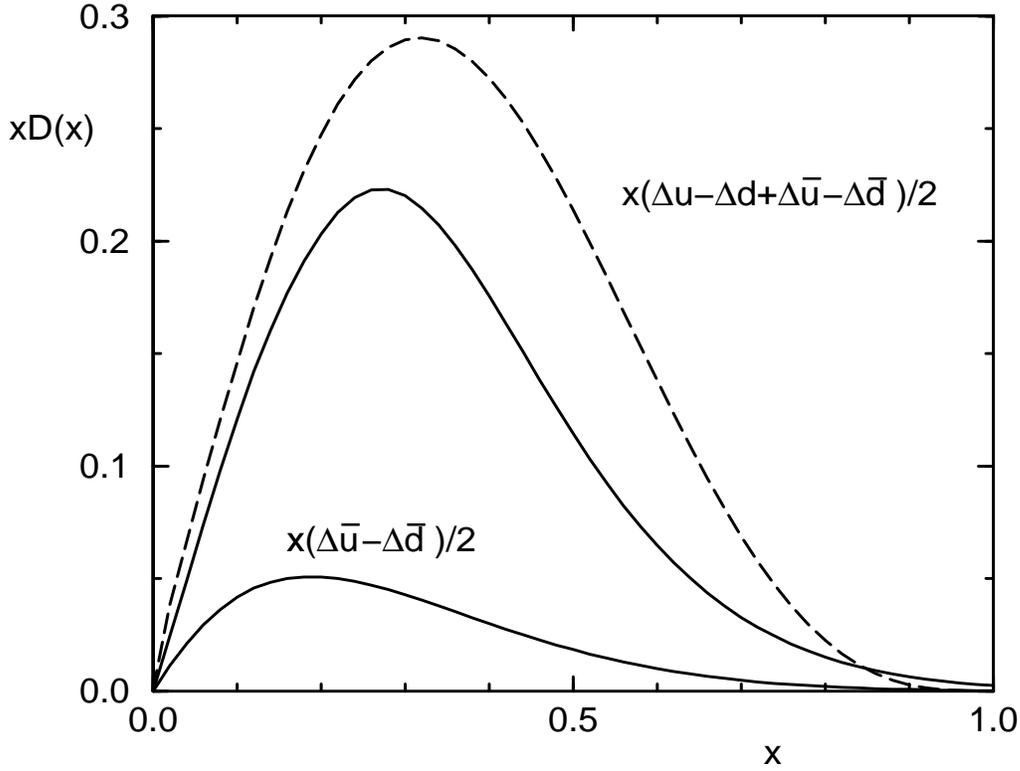}
\caption[]{{\em Solid lines}: The isovector polarized (longitud.) total 
distribution (quarks plus antiquarks) and antiquark distribution computed 
in the large--$N_c$ limit \cite{DPPPW96,DPPPW97}. {\em Dashed lines}: 
NLO--parametrization of ref.\cite{GRSV96}. The polarized antiquark 
distribution is assumed zero in the fit.}
\label{fig_2}
\end{figure}
\par
Results for the isosinglet--unpolarized and isovector--polarized
distributions are shown in Fig.1 and 2, where we compare with the
parametrizations of Gl\"uck, Reya {\em et al}. \cite{GRV95,GRSV96}.
In the calculations reported here we have chosen a simple UV 
regularization by way of Eq.(\ref{PV}) and used the standard parameters
for the effective chiral theory (see refs.\cite{DPPPW96,DPPPW97} for details);
we have not attempted to make a best fit.
Fig.1 shows the isosinglet unpolarized valence and antiquark DF.
We note that the antiquark distribution obtained 
in our approach is of the ``valence--like'' ({\em i.e.}, non-singular 
at small $x$) form assumed which was assumed in the fits of ref.\cite{GRV95}. 
Fig.2 shows the isovector polarized total distribution 
(quarks plus antiquarks) as well as the antiquark distribution,
which was set to zero in the fit of ref.\cite{GRSV96}.
That the calculated total distribution is systematically smaller than 
the fit is due to the fact that the isovector axial coupling, $g_A^{(3)}$,
is underestimated by the large--$N_c$ limit; the $1/N_c$--corrections
in this channel are large and have been computed \cite{Review}.
In Figs.1 and 2 we compare the calculated DF's with the parametrizations
at face value, without taking into account evolution; 
the normalization point of the calculated DF's may actually be 
lower than that of the parametrizations.
\par
We stress that we are computing the twist--2 parton distributions at 
a low normalization point, not the structure functions (cross sections) at 
low $q^2$, which are, in principle, directly measurable. The latter are
affected by higher--twist (power) corrections, which become large at
low $q^2$. Our calculated distributions should be used as input for
perturbative evolution, starting with a scale below the cutoff,
$\bar\rho^{-1} \simeq 600 \, {\rm MeV}$. A direct comparison
with the data can be performed only after perturbative evolution to 
sufficiently large $q^2$. 
\par
Results of a calculation of the ``small'' DF's in the large--$N_c$ limit, 
the isovector unpolarized and isosinglet polarized, have been reported 
by Wakamatsu and Kubota \cite{waka}.
\par
Recently also the off-forward quark DF's have been computed in the
chiral soliton model \cite{PPPBGW97}. These generalized DF's are defined as
nonforward matrix elements of light--ray operators of the type of 
Eq.(\ref{nonlocal}) (non-zero momentum transfer). It was found that
the large--$N_c$ approach predicts a strong dependence of the isosinglet 
off-forward distribution on the longitudinal momentum transfer.
\par
\underline{Summary.} 
To summarize, we have shown that the large--$N_c$ picture of the 
nucleon as a soliton of the effective chiral theory is able to 
explain the essential features of the quark and antiquark DF's at a low
normalization point. As we have shown, the success of this approach
is essentially due to two circumstances:
\begin{itemize}
\item 
the parametric smallness of the ratio of the dynamical quark mass to 
the cutoff of the effective theory, $M\bar\rho$ (related to 
the small packing fraction of the instanton medium), which allows
to identify the twist--2 QCD (anti--) quark DF's with the 
distributions computed in the effective theory;
\item
the field--theoretic character of this description, which ensures
that the DF's satisfy all general requirements such as positivity, 
normalization {\em etc.}
\end{itemize}
An important direction for future work is the refinement of this approach
to a level that allows one to take into account effects of higher order 
in $(M\bar\rho )^2$, {\em i.e.}, to resolve the structure of the 
``constituent'' quark. This can be done in the framework of the instanton 
vacuum, which allows to derive the effective chiral action by explicit
integration over non-perturbative gluon degrees of freedom.
In particular, at order $(M\bar\rho )^2$ the gluon distribution
will start to appear. Also, this will allow to determine the 
normalization point of the calculated DF's more accurately.


\begin{thebibliography}{99}
%
\bibitem{DPPPW96} D.I.\ Diakonov, V.Yu.\ Petrov, P.V.\ Pobylitsa,
M.V.\ Polyakov and C. Weiss, Nucl.\ Phys.\ {\bf B 480} (1996) 341.
%
\bibitem{DPPPW97} D.I.\ Diakonov, V.Yu.\ Petrov, P.V.\ Pobylitsa,
M.V.\ Polyakov and C. Weiss, Phys.\ Rev.\ {\bf D 56} (1997) 4069.
%
\bibitem{PP96} P.V.\ Pobylitsa and M.V.\ Polyakov, 
Phys.\ Lett.\ {\bf B 389} (1996) 350.
%
\bibitem{MRS95}
For a review, see: A.D.\ Martin, Acta Phys.\ Pol.\ {\bf 27} (1996) 1287; \\
A.D.\ Martin, R.G.\ Roberts, and W.J.\ Stirling, 
Phys.\ Lett.\ {\bf B 354} (1995) 155; Phys.\ Rev.\ {\bf D 50} (1994) 6734.
%
\bibitem{CTEQ95}
The CTEQ collaboration: H.L.\ Lai {\em et al.}, Phys.\ Rev.\ {\bf D 55}
(1997) 1280; Phys.\ Rev.\ {\bf D 51} (1995) 4763.
%
\bibitem{GRV95}
M. Gl\"uck, E. Reya, and A. Vogt, Z. Phys.\ {\bf C 67} (1995) 433.
%
\bibitem{GRSV96}
M. Gl\"uck, E. Reya, M. Stratmann, and W. Vogelsang, Phys.\ Rev.\
{\bf D 53} (1996) 4775.
%
\bibitem{SchSh96} T. Sch\"afer and E.V.\ Shuryak, preprint 
DOE-ER-40561-293 (1996), {\tt hep-ph/9610451}, to appear in Rev.\ Mod.\ Phys.
%
\bibitem{D96_Varenna} D.I.\ Diakonov, Talk given at the International 
School of Physics, ``Enrico Fermi'', Course 80: Selected Topics in 
Nonperturbative QCD, Varenna, Italy, Jun.\ 27 -- Jul.\ 7, 1995, 
hep-ph/9602375. 
%
\bibitem{DP84}
D. Diakonov and V. Petrov, Nucl.\ Phys.\ {\bf B 245} (1984) 259.
%
\bibitem{DP86}
D. Diakonov and V. Petrov, Nucl.\ Phys.\ {\bf B 272} (1986) 457;
LNPI preprint LNPI-1153 (1986), published in: Hadron matter under extreme
conditions, Naukova Dumka, Kiev (1986), p.192.
%
\bibitem{lattice} 
M.-C. Chu, J. Grandy, S. Huang and J. Negele, Phys.\ Rev.\ Lett.\
{\bf 70} (1993) 225; Phys.\ Rev.\ {\bf D 49} (1994) 6039; \\
T. DeGrand, A. Hasenfratz and T.G. Kovacs, Colorado Univ.\ preprint 
COLO-HEP-383 (1997), {\tt hep-lat/9705009}; \\ 
Ph.\ de Forcrand, M. Garcia Perez and I.--O.\ Stamatescu, 
Nucl.\ Phys.\ {\bf B 499} (1997) 409.
%
\bibitem{Witten}
E. Witten, Nucl.\ Phys.\ {\bf B 223} (1983) 433.
%
\bibitem{ANW}
G. Adkins, C. Nappi and E. Witten, Nucl.\ Phys.\ {\bf B 228} (1983) 552.
%
\bibitem{DPP88}
D. Diakonov and V. Petrov, Sov.\ Phys.\ JETP
Lett.\ {\bf 43} (1986) 57;  \\
D. Diakonov, V. Petrov and P. Pobylitsa, Nucl.\ Phys.\ 
{\bf B 306} (1988) 809.
%
\bibitem{Review} For a review, see: Ch.V.\ Christov {\em et al.},
Prog.\ Part.\ Nucl.\ Phys.\ {\bf 37} (1996) 91.
%
\bibitem{DE}
D. Diakonov and M. Eides, Sov.\ Phys.\ JETP Lett.\ {\bf 38} (1983) 433.
%
\bibitem{DSW}
A. Dhar, R. Shankar and S. Wadia, Phys.\ Rev.\ {\bf D 31} (1984) 3256.
%
\bibitem{DPW95} D.I.\ Diakonov, M.V.\ Polyakov and C.\ Weiss,
Nucl.\ Phys.\ {\bf B 461} (1996) 539.
%
\bibitem{Kochelev} N.I.\ Kochelev, in: Proceedings of the Workshop ``Deep
Inelastic Scattering off Polarized Targets: Theory Meets Experiment'',
DESY--Zeuthen,  Sep.\ 1--5, 1997, {\tt hep-ph/9711226};
Talk given at the 5th International Workshop on Deep Inelastic Scattering 
and QCD (DIS 97), Chicago, IL, {\tt hep-ph/9707418}. 
%
\bibitem{BPW97} J.\ Balla, M.V.\ Polyakov and C. Weiss, 
Bochum University preprint RUB-TPII-6/97, {\tt hep-ph/9707515}, 
Nucl.\ Phys.\ {\bf B}, in press.
%
\bibitem{PW97} M.V.\ Polyakov and C. Weiss, in: Proceedings of the 37th 
Cracow School of Theoretical Physics: Dynamics of Strong Interactions, 
Zakopane, Poland, May 30 --  Jun. 10, 1997, Bochum University preprint 
RUB-TPII-9/97, {\tt hep-ph/9709436}.
%
\bibitem{Feynman} R.P.\ Feynman, in: Photon--Hadron Interactions,
Benjamin, 1972.
%
\bibitem{Collins-Soper-82}
J.C.\ Collins and D.E.\ Soper, Nucl.\ Phys.\ {\bf B 194} (1982) 445.
%
\bibitem{Jaffe95} R.L.\ Jaffe, Talk given at Ettore
Majorana International School of Nucleon Structure: 1st Course: 
The Spin Structure of the Nucleon,
Erice, Italy, Aug 3--10, 1995, MIT preprint MIT-CTP-2506, 
{\tt hep-ph/9602236}. 
%
\bibitem{waka} M. Wakamatsu and T. Kubota, {\tt hep-ph/9707500}.
%
\bibitem{PPPBGW97} V.Yu.\ Petrov, P.V.\ Pobylitsa, M.V.\ Polyakov, 
I. B\"ornig, K. Goeke, and C. Weiss, Bochum University preprint 
RUB-TPII-8/97, {\tt hep-ph/9710270}.
%
\end{thebibliography}
\end{document}